\documentclass[aip,reprint]{revtex4-1}

\usepackage{graphicx}
\usepackage{dcolumn}
\usepackage{bm}
\usepackage[utf8]{inputenc}
\usepackage[T1]{fontenc}
\usepackage{mathptmx}
\usepackage{etoolbox}
\usepackage{subfiles} 
\usepackage{multirow}
\usepackage{makecell}
\usepackage[table]{xcolor}
\usepackage{booktabs}
\usepackage{longtable}
\usepackage{hyperref}

\makeatother
\draft 

\begin{document}

\title{Validation of SOLPS-ITER Simulations against the TCV-X21 Reference Case} 




\author{Y. Wang}
\email[]{yinghan.wang@epfl.ch}
\affiliation{\'Ecole Polytechnique F\'ed\'erale de Lausanne (EPFL), Swiss Plasma Center (SPC), CH-1015 Lausanne, Switzerland}

\author{C. Colandrea}
\affiliation{\'Ecole Polytechnique F\'ed\'erale de Lausanne (EPFL), Swiss Plasma Center (SPC), CH-1015 Lausanne, Switzerland}

\author{D. S. Oliveira}
\affiliation{\'Ecole Polytechnique F\'ed\'erale de Lausanne (EPFL), Swiss Plasma Center (SPC), CH-1015 Lausanne, Switzerland}

\author{C. Theiler}
\affiliation{\'Ecole Polytechnique F\'ed\'erale de Lausanne (EPFL), Swiss Plasma Center (SPC), CH-1015 Lausanne, Switzerland}
\author{H. Reimerdes}
\affiliation{\'Ecole Polytechnique F\'ed\'erale de Lausanne (EPFL), Swiss Plasma Center (SPC), CH-1015 Lausanne, Switzerland}

\author{T. Body}
\affiliation{Max Planck Institute for Plasma Physics, Garching, Germany}
\affiliation{Commonwealth Fusion Systems, Devens, Massachusetts 01434, USA}

\author{D. Galassi}
\affiliation{\'Ecole Polytechnique F\'ed\'erale de Lausanne (EPFL), Swiss Plasma Center (SPC), CH-1015 Lausanne, Switzerland}

\author{L. Martinelli}
\affiliation{\'Ecole Polytechnique F\'ed\'erale de Lausanne (EPFL), Swiss Plasma Center (SPC), CH-1015 Lausanne, Switzerland}

\author{K. Lee}
\affiliation{\'Ecole Polytechnique F\'ed\'erale de Lausanne (EPFL), Swiss Plasma Center (SPC), CH-1015 Lausanne, Switzerland}

\author{TCV Team}
\altaffiliation[]{See Reimerdes \textit{et al.} 2022 Nucl. Fusion \textbf{62} 042018}



\begin{abstract}
This paper presents a quantitative validation of SOLPS-ITER simulations against the TCV-X21 reference case and provides insights into the neutral dynamics and ionization source distribution in this scenario. TCV-X21 is a well-diagnosed diverted L-mode sheath-limited plasma scenario in both toroidal field directions, designed specifically for the validation of turbulence codes [D.S. Oliveira, T. Body, et al 2022 Nucl. Fusion \textbf{62} 096001]. Despite the optimization to reduce the impact of the neutral dynamics in this scenario, the absence of neutrals in previous turbulence simulations of TCV-X21 was identified as a possible explanation for the observed disagreements with the experimental data in the divertor region. This motivates the present study with SOLPS-ITER that includes kinetic neutral dynamics via EIRENE. Five new observables are added to the extensive, publicly available TCV-X21 dataset. These are three deuterium Balmer lines in the divertor and neutral pressure measurements in the common and private flux regions. The quantitative agreement metric used in the validation is combined with the conjugate gradient method to approach the SOLPS-ITER input parameters that return the best overall agreement with the experiment. A proof-of-principle test of this method results in a modest improvement in the level-of-agreement; the shortcomings impacting the result and how to improve the methodology are discussed. Alternatively, a scan of the particle and heat diffusion coefficients shows an improvement of 10.4\% beyond the agreement level achieved by the gradient method. The result is found for an increased transport coefficient compared to what is usually used for TCV L-mode plasmas, suggesting the need for accurate self-consistent turbulence models for predictive boundary simulations. The simulations indicate that  $\sim65\%$ of the total ionization occurs in the SOL, motivating the inclusion of neutrals in future turbulence simulations on the path towards improved agreement with the experiment.

\end{abstract}

\maketitle 

\section{\label{sec:Intro}Introduction}

The power exhaust problem is one of the key challenges faced by the magnetic confinement fusion community. Significant progress has been achieved with the introduction of diverted magnetic geometries such as the single-null configuration planned for ITER~\cite{loarte_chapter_2007}. In a single-null plasma, a poloidal magnetic null (X-point) is introduced in the plasma boundary, localizing the hot core plasma some distance away from the vacuum vessel wall and directing the "open" magnetic field lines and thus the majority of the heat flux to specially designed plates called divertor targets.  
However, it will still be challenging to keep the heat load deposited onto the target plates of future devices such as ITER~\cite{Pitts_2019} and SPARC~\cite{kuang_divertor_2020}, and even more so of a DEMO~\cite{reimerdes_assessment_2020} within technological limits. To address this issue, it is necessary to have reliable modeling of the transport and the underlying physical process governing the "open" field line region called the scrape-off layer (SOL). 
   
The investigation of the plasma dynamics in the SOL is difficult due to the limited diagnostic access and the complexity of the most complete theoretical models. Validation of numerical simulations against experimental observables stands as a suitable methodology to test and improve the capabilities of the current models~\cite{Greenwald_verification_validation_2010,ricci_Method_validation_2011}. In this approach, the largest possible number of experimental observables are quantitatively compared to the simulation results, and the overall level of agreement is quantified by an agreement metric, i.e. a single numerical value denoted by $\chi$. Such a validation provides a robust framework to assess the current agreement, effect of model improvements and, when agreement is judged satisfactory, provides confidence in the code outputs and their predictive capabilities.

The modeling of the tokamak boundary plasma can be carried out by edge plasma transport codes such as SOLPS-ITER~\cite{wensing_solps-iter_2021}, SOLEDGE2D~\cite{bufferand_numerical_2015}, UEDGE \cite{Christen_2017} or edge turbulence codes, as in the recent multi-code validation involving GBS~\cite{giacomin_gbs_2022}, GRILLIX~\cite{stegmeir_global_2019}, and TOKAM3X~\cite{tamain_tokam3x_2016}, which presented the first full tokamak size edge turbulence simulations~\cite{sales_de_oliveira_validation_2022} of a diverted single-null plasma in the Tokamak \`a Configuration Variable (TCV)~\cite{reimerdes_overview_2022}. The scenario simulated by the turbulent edge codes in this study, referred to as TCV-X21 reference case, is a L-mode plasma in sheath-limited condition to minimize the effect of the neutral dynamics in the divertor volume, as these effects were not included in the turbulence simulations. Instead, the ionization source was prescribed as an input and assumed to be localized in the outer region of the confined plasma. However, a non-negligible effect of neutrals was suggested as a possible cause of the relatively low level of agreement in the divertor region and at the divertor targets in the simulation-experiment comparison~\cite{sales_de_oliveira_validation_2022}.   

In this work, we validate the SOLPS-ITER code against the TCV-X21 scenario, as a first step to understand the role of the neutrals in this case, with kinetic neutrals simulated with EIRENE. For this purpose we extend the publicly available~\cite{sales_de_oliveira_validation_2022} TCV-X21 dataset (Tab.~\ref{tab:observables}) with two new observables for the neutral dynamics: 
Balmer line intensities measured by the Divertor Spectroscopy System (DSS)~\cite{martinelli_implementation_2022} and neutral pressure measurements from the Baratron gauges~\cite{theiler_results_2017}. We conduct a similar quantitative validation as in Ref.~\onlinecite{sales_de_oliveira_validation_2022}, using the methodology of Ref.~\onlinecite{ricci_approaching_2015} (briefly reviewed in Sec.~\ref{sec:Validation}).
With this, we expect to help guide future edge turbulence simulations of the TCV-X21 reference case including neutral dynamics. We also provide proof-of-principle tests of using different approaches to determine the SOLPS-ITER input parameters that result in the best simulation-experimental agreement, in particular, the conjugate gradient method applied on the validation metric.

The paper is organized as follows: the TCV-X21 experiment reference case and the associated dataset are introduced and extended in Sec.~\ref{sec:Experiment}. The SOLPS-ITER simulations are described in Sec.~\ref{sec:SOLPS-ITER}. Then in Sec.~\ref{sec:Validation}, we present the qualitative and quantitative validation results for simulations with the standard up-stream matching approach and the  systematic methodology to determine the input parameters based on the quantitative validation result. In Sec.~\ref{sec:Discussion}, we analyze and discuss the results of the validation and the effect of neutrals in the TCV-X21 scenario. Finally, the conclusions are presented in Sec.~\ref{sec:Conclusion}.
\section{\label{sec:Experiment} TCV-X21 Experimental dataset and its extension}

\begin{figure}[htbp]
\includegraphics[width= 0.48\textwidth ]{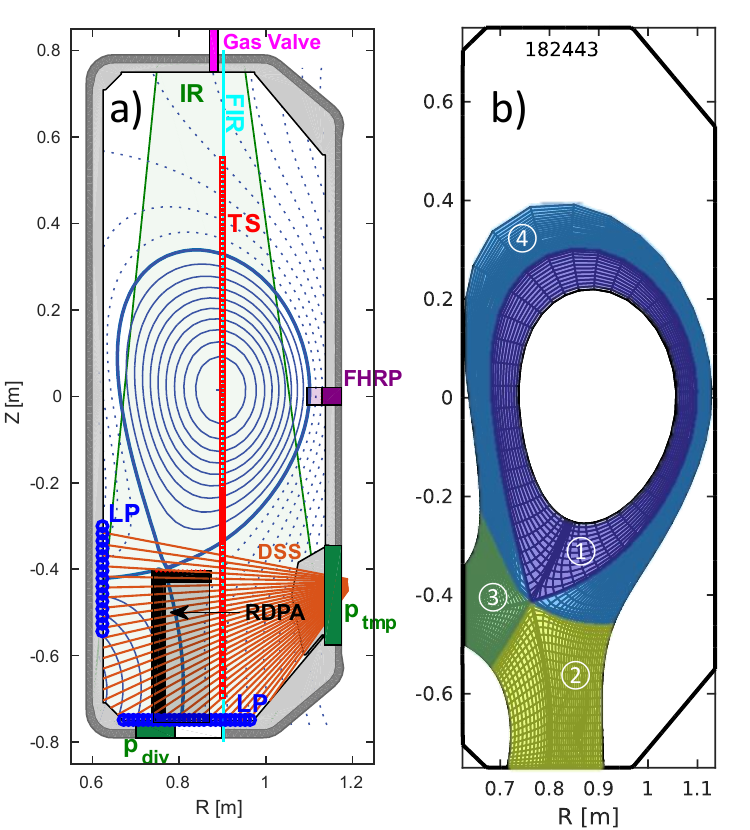}
\caption{ \label{fig:ExpSimu_polview}(a) Poloidal view of the magnetic reconstruction (dark blue lines) of the TCV-X21 reference case from LIUQE reconstruction~\cite{moret_tokamak_LIUQE_2015}, also shown are: the Langmuir probes (LP, blue circles), the Reciprocating Divertor Probe Array (RDPA, black L-shaped structure) and its covered area (transparent black rectangle), the Thomson Scattering system (TS, red square array), the Fast Horizontal Reciprocating Probe (FHRP, purple rectangle) and its covered area (transparent purple), the Far InfraRed interferometer (FIR, vertical cyan line), the area of sight of the Vertical Infrared Camera (IR, green transparent patched area), the position of the top gas fueling valve (magenta rectangle), the ports of the baratron pressure gauges (dark green boxes on the wall), and the DSS viewing chords (radial orange lines in the divertor). (b) Computational grid used for the SOLPS ITER plasma model. Regions marked by different colors are \raisebox{.5pt}{\textcircled{\raisebox{-.9pt} {1}}}: the core; \raisebox{.5pt}{\textcircled{\raisebox{-.9pt} {2}}}: the low-field-side (LFS) divertor region; \raisebox{.5pt}{\textcircled{\raisebox{-.9pt} {3}}}: the high-field-side (HFS) divertor region; and \raisebox{.5pt}{\textcircled{\raisebox{-.9pt} {4}}}: other region of the SOL.
}
\end{figure}

In this work we use the experimental dataset of the TCV-X21  reference case which is publicly available at \url{https://github.com/SPCData/TCV-X21}. This scenario is a lower single null L-mode Ohmic plasma in the TCV tokamak~\cite{reimerdes_overview_2022} with a toroidal field of $B_{\phi, axis}\simeq 0.95\mathrm{T} $, a plasma current of $I_{p}\simeq 165\mathrm{kA} $ and an electron line-average density $\langle n_e \rangle \sim 2.5\times 10^{19} \mathrm{m^{-3}}$ measured by the far infrared camera (FIR, Fig.~\ref{fig:ExpSimu_polview}(a), vertical cyan line). TCV-X21 includes data in both toroidal field directions to study the effect of drifts, with the convention that "forward" (Forw) denotes the field direction where the ion $\nabla B$ drift ($\mathbf{B} \times \nabla B$) points downwards, from the core towards the X-point and "reversed" (Rev) when it points upwards~\cite{sales_de_oliveira_validation_2022}. 

Tab.~\ref{tab:observables} lists the diagnostics, observables and their respective hierarchies (used in the validation metric) of the TCV-X21 dataset. Fig.~\ref{fig:ExpSimu_polview}(a) shows the position of the listed diagnostics. The TCV-X21 dataset includes mean and fluctuation profiles of observables covering the divertor targets and volume, the divertor entrance, and the outboard midplane. In this work, we only consider the mean profiles because SOLPS-ITER does not predict fluctuation quantities. Detailed information about the observables and diagnostics can be found in Ref.~\onlinecite{sales_de_oliveira_validation_2022}. In this work we add new observables to the dataset, namely the divertor neutral pressure measurements and deuterium Balmer lines, which will be introduced in the following subsections. 

\begin{table}[htbp]
    \centering
    \caption{\textbf{Observables and comparison hierarchies for validation.} Observables constituting the TCV-X21 dataset but not used in the present validation are shown in gray. The hierarchy weighting, $H_j$, is a constant value depending on the number of assumptions and/or models used to obtain the observable, $H_j=[h_{sim}+h_{exp}-1]^{-1}$, see Sec.~\ref{sec:Validation} and Appendix~\ref{appendix:vali_method}. The data from the Divertor Spectroscopy System (DSS) and the Baratron gauges (BAR) are newly added into the TCV-X21 dataset.} 
    \begin{tabular}{ccccc} 
    \hline
    \multirow{2}{*} {Diagnostic} & \multirow{2}{*} { Observable } & \multicolumn{3}{c} { Hierarchy }\\ 
    & & $h_{Exp}$ & $h_{Sim}$ & $H_{j}$ \\
    \hline
    
    \multirow{5}{*}{\makecell[c]{Wall Langmuir Probes (LP) \\at the low-field-side and\\ high-field-side targets }}
    & $n_e$, $T_{e}$, $V_{pl}$  & 2 & 1 & 1 / 2 \\
    ~ & $J_{\text {sat }}$, {\color{gray}$\sigma (J_{\text{sat}})$} & 1 & 2 & 1 / 2 \\
    ~ & {\color{gray}$\mathrm{skew}(J_{\text{sat}})$}, {\color{gray}$\mathrm{kurt}(J_{\text{sat}})$} & 1 & 2 & 1 / 2\\
    ~ & $J_{\parallel}$, {\color{gray}$\sigma (J_{\parallel})$} & 1 & 1 & 1 \\
    ~ & $V_{fl}$, {\color{gray}$\sigma (V_{fl})$} & 1 & 2 & 1 / 2\\
    \hline
    
    {\makecell[c]{Infrared camera (IR) \\ for low-field-side target }}
    & $q_\parallel$ & 2 & 2 & 1 / 3 \\
    \hline
    
    \multirow{5}{*}{\makecell[c]{Reciprocating divertor \\probe array (RDPA) \\for divertor volume}}
    & $n_e$, $T_{e}$, $V_{pl}$ & 2 & 1 & 1 / 2 \\
    ~ & $M_{\parallel}$ & 2 & 2 & 1 / 3 \\
    ~ & $J_{\text{sat}}$, {\color{gray}$\sigma (J_{\text{sat}})$} & 1 & 2 & 1 / 2 \\
    ~ & {\color{gray}$\mathrm{skew}(J_{\text{sat}})$}, {\color{gray}$\mathrm{kurt}(J_{\text{sat}})$} & 1 & 2 & 1 / 2\\
    ~ & $V_{fl}$, {\color{gray}$\sigma (V_{fl})$} & 1 & 2 & 1 / 2\\
    \hline
    
    {\makecell[c]{Thomson scattering (TS) \\ for divertor entrance }}
    & $n$, $T_e$ & 2 & 1 & 1 / 2 \\
    \hline
    
    \multirow{5}{*}{\makecell[c]{Fast horizontally-\\
    reciprocating probe (FHRP) \\for outboard midplane}}
    & $n_e$, $T_{e}$, $V_{pl}$ & 2 & 1 & 1 / 2 \\
    ~ & $M_{\parallel}$ & 2 & 2 & 1 / 3 \\
    ~ & $J_{\text {sat }}$, {\color{gray}$\sigma (J_{\text{sat}})$} & 1 & 2 & 1 / 2 \\
    ~ & {\color{gray}$\mathrm{skew}(J_{\text{sat}})$}, {\color{gray}$\mathrm{kurt}(J_{\text{sat}})$} & 1 & 2 & 1 / 2\\
    ~ & $V_{fl}$, {\color{gray}$\sigma (V_{fl})$} & 1 & 2 & 1 / 2\\
    \hline
    
    {\makecell[c]{Divertor spectroscopy \\ system (DSS) }}
    & $D_{5\rightarrow2}$, $D_{6\rightarrow2}$, $D_{7\rightarrow2}$ & 2 & 3 & 1 / 4 \\
    \hline
    
    \multirow{2}{*}{\makecell[c]{Baratron gauges (BAR) }}
    & $p_\textrm{tmp}$ & 2 & 2 & 1 / 3 \\
    ~ & $p_\textrm{div}$ & 1 & 2 & 1 / 2 \\
    \hline
    \end{tabular}
    
    \label{tab:observables}
\end{table}


\subsection{Baratron Pressure Gauges}
The baratron pressure gauges (BAR) considered here provide measurements of neutral pressure at the TCV floor ($p_{\text{div}}$) and in the turbo pump duct ($p_{\text{tmp}}$ - see Fig.~\ref{fig:ExpSimu_polview}(a)). The gauges are installed at the end of extension tubes to protect them from the magnetic field of the tokamak. Therefore, we need a model to relate the measurement to the in-vessel pressure~\cite{theiler_results_2017}.  
The energetic atomic and molecular divertor neutrals flowing into the tube undergo thermalization through collisions with the walls and the atoms recombine to form molecules. At the end of the tube, after several bends, the pressure is determined solely by the molecular density at wall temperature, $T_\mathrm{wall}=300\mathrm{K}$. For the experiment-simulation comparision, we use the 0D model discussed in Ref.~\onlinecite{wensing_solps-iter_2019} and Ref.~\onlinecite{niemczewski_neutral_1995} to determine $p_{\text{div}}$ and $p_{\text{tmp}}$ from SOLPS-ITER outputs. 
To compensate for the pressure drop in $p_{\text{tmp}}$ induced by the turbo pump, the experimental data of $p_{\text{tmp}}$ is multiplied by a factor of $1.5$~\cite{fevrier_divertor_2021}. 
The measured pressure is averaged over $0.8\mathrm{s}$ of the flattop phase of several, repeat discharges of the TCV-X21 scenario. 

The main source of uncertainty of the $p_{\text{div}}$ and $p_{\text{tmp}}$ measurements is $\Delta e_{rep}$, the uncertainty related to reproducibility~\cite{sales_de_oliveira_validation_2022}, estimated from repeat discharges.

\subsection{Divertor Spectroscopy System}

The Divertor Spectroscopy System (DSS) installed in TCV (Fig.~\ref{fig:ExpSimu_polview}(a)) provides measurements of the line-integrated visible radiation at different wavelengths, corresponding to different atomic processes~\cite{verhaegh_spectroscopic_2017}. The DSS system has 30 chords along which it is possible to measure the wavelength spectra with a spectral resolution of up to $0.02 \mathrm{\ nm}$~\cite{martinelli_implementation_2022}. 
The measurement is a line integral of the emission along a given chord.
For each discharge, the time traces are averaged over $0.22\mathrm{s}$, providing $D_{n\rightarrow 2}$ as a function of the chord number. The final average emission profile is obtained by averaging the profiles from different shots. In this work, we consider three deuterium Balmer lines, ${D}_{5\rightarrow 2}$, ${D}_{6\rightarrow 2}$, and ${D}_{7\rightarrow 2}$.

The main sources of uncertainty of the line brightnesses are $\Delta e_{rep}$, which is estimated comparing different shots, and $\Delta e_{dia}$, the uncertainty due to inherent characteristics of the diagnostics, which is estimated as $15\%$ of the measured intensity.

\section{\label{sec:SOLPS-ITER}SOLPS-ITER Simulations }

SOLPS-ITER (Scrape-Off Layer Plasma Simulation-ITER) is composed of the transport code B2.5  that solves the Braginskii multi-fluid equations, and the kinetic neutral Monte Carlo solver EIRENE~\cite{wiesen_new_2015}. EIRENE is coupled self consistently with B2.5 to calculate the sources and sinks due to plasma-neutral interactions. The simulations in this work consider a multi-component plasma, including carbon impurities and kinetic neutrals and their main reactions in the plasma~\cite{Wensing:285066}. Drifts are not included in these simulations, since convergence in low density, high temperature plasmas could not be achieved so far for SOLPS-ITER simulations of TCV plasmas. Unlike in the previous turbulence validation, \textit{ad-hoc} diffusion coefficients for cross-field heat and particle transport are used. The simulations presented here are a good testbed for the role of the neutrals in the TCV-X21 scenario, in particular, enabling the investigation of the distribution of the ionization profile across the SOL. The absence of drifts in these simulations is a limitation, but may help disentangle the origin of the flows in the divertor, i.e., whether drift or transport-driven~\cite{smick_transport_2013}.  
 
 Fig.~\ref{fig:ExpSimu_polview}(b) shows the computational grid used in this work. The radial particle diffusion and heat conduction is described using spatially uniform anomalous diffusion coefficients. For the simulations presented in  Sec.~\ref{subsec:Vali_Standard}, we choose a particle diffusivity of $D_{\perp} = 0.2 \ \mathrm{m^2/s}$ and a thermal diffusivity of $\chi_{e,\perp}= 1.0 \ \mathrm{m^2/s}$ and $\chi_{i,\perp}= 1.0\ \mathrm{m^2/s}$, which were found in previous works to result in a good upstream match for TCV L-mode plasmas~\cite{wensing_solps-iter_2021}.  A deuterium gas puff rate of $\Gamma_\mathrm{D_2} = 6.8\times 10^{19} \mathrm{/s}$ is used for a close upstream density match, determined after a scan of the gas puff $\Gamma_\mathrm{D_2} = \{4.5, 5.6, 6.8, 7.2, 7.6, 8.0, 8.4\}\times 10^{19} \mathrm{/s}$. 
 The chemical sputtering coefficient of carbon impurities on the wall is assumed as  $Y_{\text{chem}}=3.5\%$, and the particle recycling coefficient is set to be $R=0.99$. 

At the divertor targets, sheath Bohm boundary conditions are applied.  
At the core boundary, power transferred from the core to the edge is set to be comparable to the experimental value ($160 \ \mathrm{kW}$). Neutrals crossing the core boundary are returned as fully ionized particles. More details about the boundary conditions used can be found in Ref.~\onlinecite{Wensing:285066}.

In Sec.~\ref{subsec:Gradient_method}, an iterative method using the overall-agreement metric and the conjugate gradient is used to adapt the input parameters $\Gamma_\mathrm{D_2}$,  $D_{\perp}$ and $\chi_{e,\perp}$ to achieve a better overall agreement. 
In Sec.~\ref{subsec:grid_scan}, $D_\perp$ and $\chi_\mathrm{e, \perp}$ are scanned to find a better overall agreement.

\section{\label{sec:Validation}Validation Results}

The validation of the SOLPS-ITER simulations in this work follows the same methodology as used in Ref.~\onlinecite{sales_de_oliveira_validation_2022}. 
The details of the mathematical model can be found in Ref.~\onlinecite{ricci_approaching_2015}, and the basic concept of this methodology is summarized as follows:
The level of agreement between simulation and experiment is evaluated using a large set of observables and is quantified by the \textit{overall agreement metric} $\chi$. $\chi=0$ and $\chi=1$, means, respectively, perfect agreement and complete disagreement.
The fundamental quantity used to calculate $\chi$ is the \textit{normalised simulation-experimental distance} $d_j$, which, for each observable $j$ is defined as:
\begin{equation}\label{eq:Vali_d_J}
    d_j = \left[
    \frac{1}{N_j}\sum_{i=1}^{N_j}
    \frac{(s_{j,i} - e_{j,i})^2}
    {\Delta s_{j,i}^2 + \Delta e_{j,i}^2}
    \right]^{1/2}
\end{equation}
where $e_{j,i}$, $\Delta e_{j,i}$,  $s_{j,i}$, and $\Delta s_{j,i}$ denote, respectively, the experimental values, their uncertainties, the simulation values, and their uncertainties, defined on a series of discrete data points $i\in\{1, 2,  ..., N_j \}$.
 $d_j\rightarrow0$ means a perfect agreement between simulation and experiment for the observable $j$. 
Due to the difficulty to provide a rigorous estimate of the simulation uncertainty, we set $\Delta s_{j,i}=0$ as in Ref.~\onlinecite{sales_de_oliveira_validation_2022}. 
Other important quantities used in the validation are: the \textit{sensitivity} $S_j$ of an observable $j$, which takes values between $0$ and $1$, approaching $1$ for very small relative uncertainties (high precision) of the experimental and simulation observables; the \textit{hierarchy weighting} $H_j$, as given in Tab.\ref{tab:observables}, is a value associated with each observable $j$ that is smaller the higher the number of model assumptions and/or measurement combinations needed to determine the observable.
Based on $d_j$ and the weighting factors $S_j$ and $H_j$, a metric $\chi$ is then evaluated to indicate the overall simulation-experiment agreement over all (or a subset of) the observables, see Appendix~\ref{appendix:vali_method} for more information.
In addition, the \textit{quality} $Q$ is evaluated, which denotes the quality of the validation. This value is higher when a higher number of more directly computed, higher precision observables are included in the validation, see  Appendix~\ref{appendix:vali_method}.

\begin{figure}[htbp]
\includegraphics[width= 0.49\textwidth ]{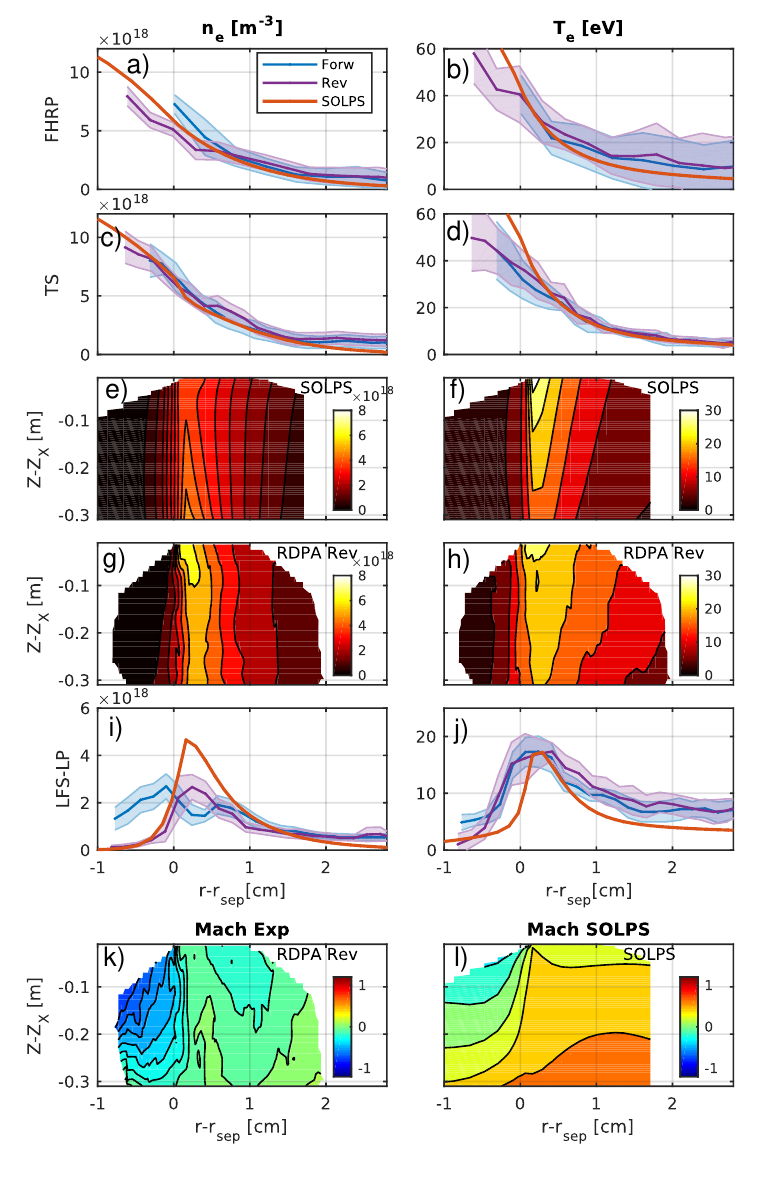}
\caption{\label{fig:Compare_182574}
\textbf{Comparison of the profiles in the standard approach of matching upstream profiles.} The 1-D experimental profiles in forward field (blue lines) and reversed field (purple lines) and in the SOLPS-ITER simulation (orange lines) for electron density (left column) and electron temperature (right column). The measurements are from the FHRP (see Fig.~\ref{fig:ExpSimu_polview}(a)) at the outboard midplane ((a) and (b)), from TS at the divertor entrance ((c) and (d)), and from LPs at the low-field-side divertor target ((i) and (j)), also shown are (e) the 2-D profile of electron density, and f) temperature of the simulation and the corresponding reversed field experiments (g) and (h) measured by RDPA in the divertor volume. (k) and (l) show the 2D Mach number measured in the experiment (reversed field) and from the SOLPS simulations.}
\end{figure}

\begin{figure*}[htbp]
\includegraphics[width= 1.0\textwidth ]{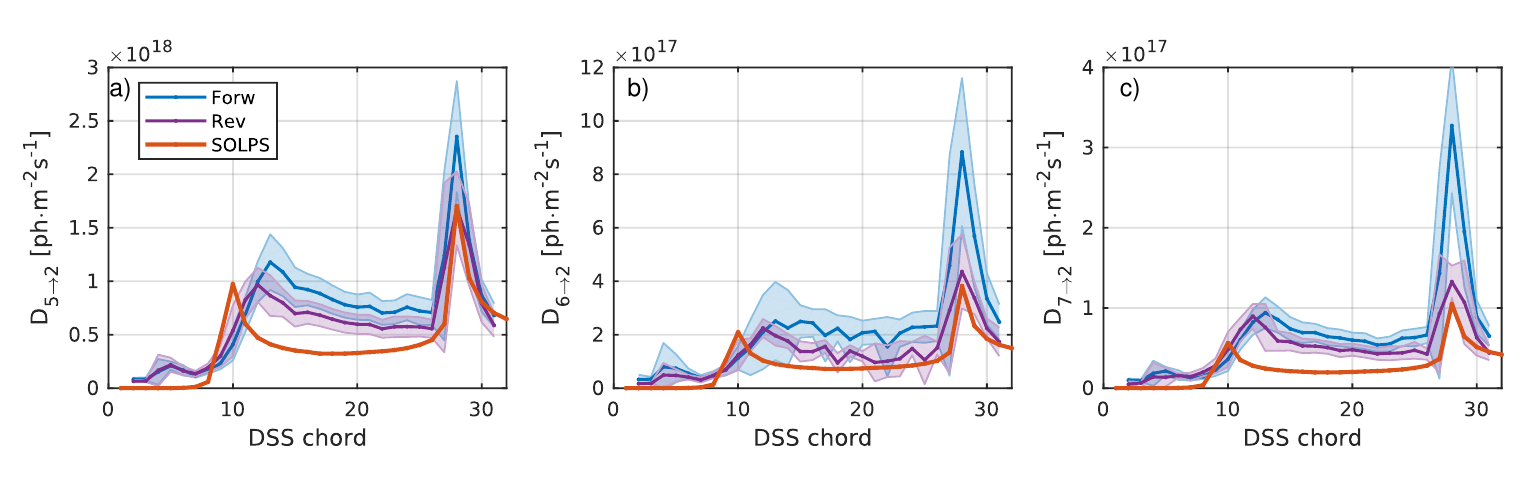}
\caption{ \label{fig:Compare_182574_DSS}
\textbf{Comparison of the DSS Balmer line profiles.} The 1-D experimental profiles of forward field (blue lines) and reversed field (purple lines) and the SOLPS-ITER simulation profiles (orange lines) for three deuterium Balmer line intensities (a) $\mathrm{D}_{5\rightarrow 2}$, (b) $\mathrm{D}_{6\rightarrow 2}$, and (c) $\mathrm{D}_{7\rightarrow 2}$ measured by DSS.}
\end{figure*}

\begin{figure*}
\includegraphics[width= 0.8\textwidth ]{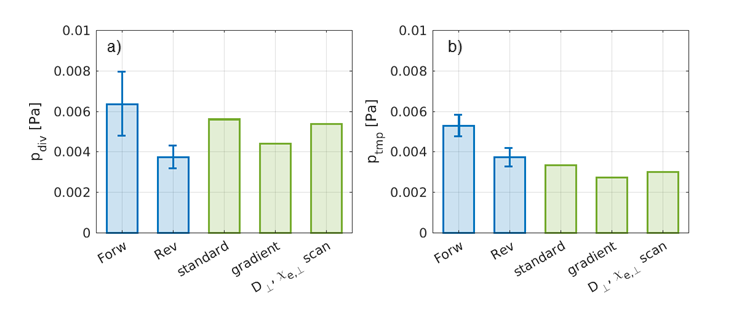}
\caption{\label{fig:Compare_182574_pn}
\textbf{Baratron neutral pressure.} The experimental baratron measurements (blue bars) and the simulation results using the synthetic diagnostic from Ref.~\onlinecite{wensing_solps-iter_2019} (green bars) at (a) the divertor floor ($p_\mathrm{div}$) and (b) the turbo pump ($p_\mathrm{tmp}$). In each subfigure we show the experimental measurements in forward (Forw) and reversed (Rev) field directions, and the simulation value in the standard approach (standard, Sec.~\ref{subsec:Vali_Standard}) and for the best agreement cases in the conjugate gradient method (gradient, Sec.~\ref{subsec:Gradient_method}) and the $D_\perp$ and $\chi_{e,\perp}$ scan ($D_\perp$, $\chi_{e,\perp}$ scan, Sec.~\ref{subsec:grid_scan}).}
\end{figure*}

\subsection{\label{subsec:Vali_Standard}Standard approach: matching upstream profiles}

The validation results for the standard approach of matching upstream profiles are given in Tab.~\ref{tab:validation_182574}, in the columns labeled Standard (Forw) and Standard (Rev). Since the present simulations do not include drifts, the results are compared to experimental data in both field directions. Some selected profiles showing simulation-experiment comparisons are given in Fig.~\ref{fig:Compare_182574}, where the radial coordinate $r-r_{\text{sep}}$ denotes the distance from the separatrix, after mapping along the magnetic flux surface to the outboard midplane.

The comparison at the outboard midplane and the divertor entrance are shown for the electron density $n_e$ and the electron temperature $T_e$ in Fig.~\ref{fig:Compare_182574}(a) to ~\ref{fig:Compare_182574}(d). An appreciable match is found as expected from the values in  Tab.~\ref{tab:validation_182574}, and since the simulations are tuned for a good upstream match. As in the simulations, the $J_{sat}$ is estimated from $n_e$ and $T_e$, its good outboard midplane agreement in Tab.~\ref{tab:validation_182574} is also expected. 
The good agreement observed for the upstream $V_{pl}$ in Tab.~\ref{tab:validation_182574} is a consequence mostly of large experimental error bars, as can be seen from the lower value of $S_j$ compared to that of $n_e$ and $T_e$. On the other hand, the floating potential $V_{fl}$ and Mach number $M_{\parallel}$ at the outboard midplane show considerable discrepancy between simulation and experiment as indicated by the $d_j$ values in Tab.~\ref{tab:validation_182574}. The value of $M_{\parallel}$ is approximately zero, different from the experiment and simulations in Ref.~\onlinecite{sales_de_oliveira_validation_2022} (not shown). This is attributed to the absence of drifts, which are the basic mechanism of the Pfirsch–Schlüter flows at the outboard midplane~\cite{pitts_parallel_2007, smick_transport_2013}.

In the divertor volume (Fig.~\ref{fig:Compare_182574}(e) to ~\ref{fig:Compare_182574}(h)), the simulation roughly reproduces the radial shape of the experimental $n_e$ and $T_e$ profiles. However, the simulated $n_e$ profile peaks at the target, while in the experiments it peaks at the X-point. For $T_e$, both simulation and experimental peaks are close to the X-point, with the simulation showing a stronger poloidal gradient.  The experimental $n_e$ and $T_e$ profiles show overall similar trends in this region in both field directions, although the density profile is shifted more towards the private flux region in forward field. 
For the Mach number in the divertor volume (Fig.~\ref{fig:Compare_182574}(k) and ~\ref{fig:Compare_182574}(l)), the SOLPS simulation predicts high values in the region $r-r_\mathrm{sep}>0$, where RDPA measured Mach numbers close to zero. 

At the low-field-side (LFS) target (Fig.~\ref{fig:Compare_182574}(i) and ~\ref{fig:Compare_182574}(j)), the simulation reproduces the peak magnitude of $T_e$, but gives a larger peak $n_e$ value. The overestimation of $n_e$ was also observed in the reversed field direction in previous SOLPS-ITER simulations of TCV L-mode plasma at $1.45 \mathrm{\ T}$~\cite{wensing_solps-iter_2021}. The simulation profile of $n_e$ and $T_e$ in Fig.~\ref{fig:Compare_182574}(i)-~\ref{fig:Compare_182574}(j) are narrower compared with the experiment in both field directions. It is worth noting that in the forward field direction, the $n_e$ experimental profile shows a double peak structure not present in the simulations. Such profile shape is usually attributed to the effect of drifts~\cite{canal_enhanced_2015}, which are not included in this simulation.

\newpage
\begin{table*}[htbp] 
\caption{\label{tab:validation_182574} 
\textbf{Quantitative validation result for SOLPS-ITER simulations.} For each observable, the normalized distance $d_j$ and the sensitivity weighting $S_j$ are listed. The composite agreement metric $\chi$ and the overall quality $Q$ are given, for each diagnostic and for the overall validation. "Standard", "Gradient", and "$D_\perp$ $\chi_\mathrm{e, \perp}$ scan" refer to the simulation with the standard upstream matching approach (Sec.~\ref{subsec:Vali_Standard}), the best agreement point in the first 1-D search of the gradient method (Sec.~\ref{subsec:Gradient_method}), and the best agreement point in the grid scan of $D_\perp$ and $\chi_e$ (Sec.~\ref{subsec:grid_scan}), respectively. In the standard approach, the simulation is validated against the two field directions in the TCV-X21 dataset~\cite{sales_de_oliveira_validation_2022}. 
}
\begin{ruledtabular}
\begin{tabular}{cccccccccc} 
 & & \multicolumn{2}{c}{Standard (Forw)}& \multicolumn{2}{c}{Standard (Rev)}& \multicolumn{2}{c}{Gradient (Rev)}& \multicolumn{2}{c}{$D_\perp$ $\chi_\mathrm{e, \perp}$ scan(Rev)}\\ 
Diagnostics & Observables & $d_j$ & $S_j$ & $d_j$ & $S_j$ & $d_j$ & $S_j$ & $d_j$ & $S_j$ \\ 
\midrule 
\multirow{7}[4]{3.0cm}{\centering Fast horizontally reciprocating probe (FHRP) for outboard midplane} & $n_e$  & \cellcolor[rgb]{ 0.67 0.86 0.43 }0.800 & 0.835 & \cellcolor[rgb]{ 0.77 0.90 0.49 }1.275 & 0.889 & \cellcolor[rgb]{ 0.67 0.86 0.43 }0.809 & 0.880 & \cellcolor[rgb]{ 0.64 0.84 0.42 }0.683 & 0.884 \\ 
 & $T_e$  & \cellcolor[rgb]{ 0.55 0.80 0.40 }0.359 & 0.733 & \cellcolor[rgb]{ 0.69 0.87 0.44 }0.907 & 0.791 & \cellcolor[rgb]{ 0.74 0.89 0.47 }1.151 & 0.799 & \cellcolor[rgb]{ 0.61 0.83 0.41 }0.567 & 0.792 \\ 
 & $J_{sat}$  & \cellcolor[rgb]{ 0.84 0.93 0.53 }1.523 & 0.905 & \cellcolor[rgb]{ 0.89 0.95 0.60 }1.845 & 0.894 & \cellcolor[rgb]{ 0.88 0.95 0.58 }1.780 & 0.890 & \cellcolor[rgb]{ 0.84 0.93 0.53 }1.559 & 0.891 \\ 
 & $V_{pl}$  & \cellcolor[rgb]{ 0.66 0.85 0.42 }0.797 & 0.695 & \cellcolor[rgb]{ 0.80 0.91 0.51 }1.386 & 0.734 & \cellcolor[rgb]{ 0.76 0.90 0.49 }1.248 & 0.743 & \cellcolor[rgb]{ 0.75 0.89 0.48 }1.179 & 0.745 \\ 
 & $V_{fl}$  & \cellcolor[rgb]{ 0.87 0.95 0.58 }1.749 & 0.703 & \cellcolor[rgb]{ 0.99 0.70 0.40 }4.217 & 0.835 & \cellcolor[rgb]{ 0.99 0.78 0.46 }3.851 & 0.846 & \cellcolor[rgb]{ 0.99 0.73 0.42 }4.091 & 0.830 \\ 
 & $M_{\parallel}$  & \cellcolor[rgb]{ 0.96 0.48 0.29 }7.295 & 0.847 & \cellcolor[rgb]{ 0.96 0.48 0.29 }6.879 & 0.876 & \cellcolor[rgb]{ 0.96 0.48 0.29 }6.881 & 0.876 & \cellcolor[rgb]{ 0.96 0.48 0.29 }6.946 & 0.875 \\ 
\cmidrule{2-10}          & ($\chi$; $Q$) & \multicolumn{2}{c}{(0.310; 2.218)}   & \multicolumn{2}{c}{(0.501; 2.363)}   & \multicolumn{2}{c}{(0.459; 2.371)}   & \multicolumn{2}{c}{(0.388; 2.362)}   \\ 
\midrule 
\multirow{7}[4]{3.0cm}{\centering Reciprocating divertor probe array (RDPA) for divertor volume} & $n_e$  & \cellcolor[rgb]{ 1.00 0.99 0.74 }2.559 & 0.871 & \cellcolor[rgb]{ 0.97 0.99 0.71 }2.335 & 0.888 & \cellcolor[rgb]{ 1.00 0.98 0.72 }2.640 & 0.881 & \cellcolor[rgb]{ 0.99 1.00 0.74 }2.467 & 0.888 \\ 
 & $T_e$  & \cellcolor[rgb]{ 0.92 0.97 0.64 }2.041 & 0.899 & \cellcolor[rgb]{ 0.90 0.96 0.61 }1.929 & 0.901 & \cellcolor[rgb]{ 0.85 0.94 0.55 }1.642 & 0.906 & \cellcolor[rgb]{ 0.75 0.89 0.48 }1.196 & 0.909 \\ 
 & $J_{sat}$  & \cellcolor[rgb]{ 0.99 0.71 0.40 }4.191 & 0.909 & \cellcolor[rgb]{ 1.00 0.81 0.49 }3.722 & 0.917 & \cellcolor[rgb]{ 1.00 0.85 0.53 }3.506 & 0.913 & \cellcolor[rgb]{ 0.99 0.71 0.40 }4.164 & 0.917 \\ 
 & $V_{pl}$  & \cellcolor[rgb]{ 1.00 0.99 0.74 }2.588 & 0.869 & \cellcolor[rgb]{ 1.00 0.80 0.48 }3.748 & 0.884 & \cellcolor[rgb]{ 1.00 0.89 0.56 }3.367 & 0.890 & \cellcolor[rgb]{ 1.00 0.92 0.62 }3.087 & 0.892 \\ 
 & $V_{fl}$  & \cellcolor[rgb]{ 0.96 0.48 0.29 }10.122 & 0.745 & \cellcolor[rgb]{ 0.96 0.48 0.29 }23.186 & 0.816 & \cellcolor[rgb]{ 0.96 0.48 0.29 }21.370 & 0.828 & \cellcolor[rgb]{ 0.96 0.48 0.29 }22.232 & 0.818 \\ 
 & $M_{\parallel}$  & \cellcolor[rgb]{ 0.96 0.48 0.29 }11.944 & 0.898 & \cellcolor[rgb]{ 0.96 0.48 0.29 }13.329 & 0.901 & \cellcolor[rgb]{ 0.96 0.48 0.29 }12.870 & 0.899 & \cellcolor[rgb]{ 0.96 0.48 0.29 }12.374 & 0.895 \\ 
\cmidrule{2-10}          & ($\chi$; $Q$) & \multicolumn{2}{c}{(0.979; 2.446)}   & \multicolumn{2}{c}{(0.966; 2.503)}   & \multicolumn{2}{c}{(0.915; 2.509)}   & \multicolumn{2}{c}{(0.829; 2.510)}   \\ 
\midrule 
\multirow{3}[4]{3.0cm}{\centering Thomson scattering (TS) for divertor entrance} & $n_e$  & \cellcolor[rgb]{ 0.67 0.86 0.43 }0.848 & 0.876 & \cellcolor[rgb]{ 0.74 0.89 0.47 }1.148 & 0.898 & \cellcolor[rgb]{ 0.80 0.91 0.51 }1.393 & 0.891 & \cellcolor[rgb]{ 0.74 0.89 0.47 }1.153 & 0.894 \\ 
 & $T_e$  & \cellcolor[rgb]{ 0.67 0.86 0.43 }0.818 & 0.893 & \cellcolor[rgb]{ 0.73 0.88 0.47 }1.095 & 0.901 & \cellcolor[rgb]{ 0.79 0.91 0.51 }1.342 & 0.905 & \cellcolor[rgb]{ 0.69 0.87 0.44 }0.908 & 0.900 \\ 
\cmidrule{2-10}          & ($\chi$; $Q$) & \multicolumn{2}{c}{(0.004; 0.884)}   & \multicolumn{2}{c}{(0.045; 0.899)}   & \multicolumn{2}{c}{(0.190; 0.898)}   & \multicolumn{2}{c}{(0.031; 0.897)}   \\ 
\midrule 
\multirow{2}[4]{3.0cm}{\centering Infrared camera (IR) for low-field-side target} & $q_\parallel$  & \cellcolor[rgb]{ 0.96 0.48 0.29 }6.350 & 0.911 & \cellcolor[rgb]{ 0.96 0.48 0.29 }6.322 & 0.941 & \cellcolor[rgb]{ 0.96 0.48 0.29 }6.437 & 0.943 & \cellcolor[rgb]{ 0.96 0.48 0.29 }6.588 & 0.943 \\ 
\cmidrule{2-10}          & ($\chi$; $Q$) & \multicolumn{2}{c}{(1.000; 0.304)}   & \multicolumn{2}{c}{(1.000; 0.314)}   & \multicolumn{2}{c}{(1.000; 0.314)}   & \multicolumn{2}{c}{(1.000; 0.314)}   \\ 
\midrule 
\multirow{7}[4]{3.0cm}{\centering Wall Langmuir probes for low-field-side target} & $n_e$  & \cellcolor[rgb]{ 1.00 0.90 0.58 }3.222 & 0.879 & \cellcolor[rgb]{ 0.92 0.97 0.64 }2.046 & 0.880 & \cellcolor[rgb]{ 0.85 0.94 0.55 }1.667 & 0.866 & \cellcolor[rgb]{ 0.91 0.96 0.62 }1.953 & 0.881 \\ 
 & $T_e$  & \cellcolor[rgb]{ 0.94 0.97 0.67 }2.170 & 0.891 & \cellcolor[rgb]{ 0.90 0.96 0.61 }1.907 & 0.866 & \cellcolor[rgb]{ 0.84 0.93 0.54 }1.599 & 0.876 & \cellcolor[rgb]{ 0.77 0.90 0.49 }1.293 & 0.880 \\ 
 & $J_{sat}$  & \cellcolor[rgb]{ 1.00 0.81 0.49 }3.688 & 0.909 & \cellcolor[rgb]{ 0.97 0.99 0.71 }2.342 & 0.908 & \cellcolor[rgb]{ 0.92 0.97 0.64 }2.066 & 0.902 & \cellcolor[rgb]{ 0.97 0.99 0.71 }2.347 & 0.909 \\ 
 & $V_{pl}$  & \cellcolor[rgb]{ 1.00 0.87 0.54 }3.441 & 0.888 & \cellcolor[rgb]{ 0.96 0.48 0.29 }4.982 & 0.887 & \cellcolor[rgb]{ 0.98 0.63 0.36 }4.483 & 0.893 & \cellcolor[rgb]{ 0.99 0.75 0.44 }4.031 & 0.897 \\ 
 & $V_{fl}$  & \cellcolor[rgb]{ 0.94 0.97 0.67 }2.184 & 0.639 & \cellcolor[rgb]{ 0.99 0.66 0.37 }4.356 & 0.732 & \cellcolor[rgb]{ 0.99 0.70 0.40 }4.237 & 0.741 & \cellcolor[rgb]{ 0.99 0.69 0.38 }4.278 & 0.733 \\ 
 & $J_{\parallel}$  & \cellcolor[rgb]{ 1.00 0.87 0.54 }3.451 & 0.746 & \cellcolor[rgb]{ 1.00 0.93 0.63 }3.015 & 0.768 & \cellcolor[rgb]{ 1.00 0.93 0.63 }3.028 & 0.767 & \cellcolor[rgb]{ 1.00 0.95 0.67 }2.903 & 0.755 \\ 
\cmidrule{2-10}          & ($\chi$; $Q$) & \multicolumn{2}{c}{(0.985; 2.849)}   & \multicolumn{2}{c}{(0.954; 2.904)}   & \multicolumn{2}{c}{(0.842; 2.906)}   & \multicolumn{2}{c}{(0.841; 2.906)}   \\ 
\midrule 
\multirow{7}[4]{3.0cm}{\centering Wall Langmuir probes for high-field-side target} & $n_e$  & \cellcolor[rgb]{ 1.00 0.89 0.56 }3.360 & 0.886 & \cellcolor[rgb]{ 0.99 0.69 0.38 }4.269 & 0.901 & \cellcolor[rgb]{ 1.00 0.93 0.63 }3.046 & 0.888 & \cellcolor[rgb]{ 1.00 0.83 0.51 }3.602 & 0.901 \\ 
 & $T_e$  & \cellcolor[rgb]{ 1.00 0.92 0.61 }3.136 & 0.922 & \cellcolor[rgb]{ 1.00 0.97 0.70 }2.750 & 0.892 & \cellcolor[rgb]{ 0.96 0.98 0.69 }2.248 & 0.899 & \cellcolor[rgb]{ 0.84 0.93 0.53 }1.531 & 0.902 \\ 
 & $J_{sat}$  & \cellcolor[rgb]{ 0.92 0.97 0.64 }2.079 & 0.878 & \cellcolor[rgb]{ 1.00 0.81 0.49 }3.702 & 0.900 & \cellcolor[rgb]{ 1.00 0.93 0.63 }3.054 & 0.892 & \cellcolor[rgb]{ 1.00 0.94 0.64 }2.992 & 0.900 \\ 
 & $V_{pl}$  & \cellcolor[rgb]{ 0.96 0.48 0.29 }5.234 & 0.915 & \cellcolor[rgb]{ 0.98 0.56 0.32 }4.756 & 0.884 & \cellcolor[rgb]{ 0.98 0.63 0.36 }4.490 & 0.890 & \cellcolor[rgb]{ 0.99 0.76 0.44 }3.972 & 0.895 \\ 
 & $V_{fl}$  & \cellcolor[rgb]{ 1.00 0.95 0.67 }2.907 & 0.673 & \cellcolor[rgb]{ 1.00 0.83 0.51 }3.583 & 0.687 & \cellcolor[rgb]{ 1.00 0.85 0.52 }3.537 & 0.696 & \cellcolor[rgb]{ 1.00 0.81 0.49 }3.730 & 0.687 \\ 
 & $J_{\parallel}$  & \cellcolor[rgb]{ 1.00 0.83 0.51 }3.617 & 0.788 & \cellcolor[rgb]{ 1.00 0.82 0.50 }3.648 & 0.781 & \cellcolor[rgb]{ 1.00 0.82 0.50 }3.643 & 0.780 & \cellcolor[rgb]{ 0.99 0.78 0.46 }3.846 & 0.777 \\ 
\cmidrule{2-10}          & ($\chi$; $Q$) & \multicolumn{2}{c}{(0.987; 2.925)}   & \multicolumn{2}{c}{(0.999; 2.913)}   & \multicolumn{2}{c}{(0.994; 2.912)}   & \multicolumn{2}{c}{(0.904; 2.920)}   \\ 
\midrule 
\multirow{4}[4]{3.0cm}{\centering Divertor spectroscopy system (DSS) for divertor volume} & $D_{5\rightarrow2}$  & \cellcolor[rgb]{ 1.00 0.87 0.54 }3.460 & 0.869 & \cellcolor[rgb]{ 1.00 0.94 0.64 }3.014 & 0.870 & \cellcolor[rgb]{ 1.00 0.90 0.58 }3.240 & 0.862 & \cellcolor[rgb]{ 1.00 0.95 0.67 }2.830 & 0.876 \\ 
 & $D_{6\rightarrow2}$  & \cellcolor[rgb]{ 0.95 0.98 0.68 }2.218 & 0.753 & \cellcolor[rgb]{ 1.00 0.98 0.72 }2.627 & 0.825 & \cellcolor[rgb]{ 1.00 0.96 0.69 }2.774 & 0.818 & \cellcolor[rgb]{ 0.99 1.00 0.74 }2.481 & 0.835 \\ 
 & $D_{7\rightarrow2}$  & \cellcolor[rgb]{ 1.00 0.87 0.54 }3.467 & 0.818 & \cellcolor[rgb]{ 1.00 0.95 0.67 }2.896 & 0.826 & \cellcolor[rgb]{ 1.00 0.93 0.63 }3.053 & 0.821 & \cellcolor[rgb]{ 1.00 0.98 0.72 }2.641 & 0.835 \\ 
\cmidrule{2-10}          & ($\chi$; $Q$) & \multicolumn{2}{c}{(0.986; 0.610)}   & \multicolumn{2}{c}{(0.997; 0.630)}   & \multicolumn{2}{c}{(0.998; 0.625)}   & \multicolumn{2}{c}{(0.993; 0.637)}   \\ 
\midrule 
\multirow{3}[4]{3.0cm}{\centering Baratron pressure gauges (BAR) for divertor volume} & $p_{\textrm{div}}$  & \cellcolor[rgb]{ 0.58 0.82 0.41 }0.469 & 0.877 & \cellcolor[rgb]{ 1.00 0.89 0.56 }3.336 & 0.941 & \cellcolor[rgb]{ 0.75 0.89 0.48 }1.193 & 0.933 & \cellcolor[rgb]{ 1.00 0.95 0.67 }2.871 & 0.940 \\ 
 & $p_{\textrm{tmp}}$  & \cellcolor[rgb]{ 1.00 0.82 0.50 }3.685 & 0.940 & \cellcolor[rgb]{ 0.67 0.86 0.43 }0.845 & 0.936 & \cellcolor[rgb]{ 0.93 0.97 0.66 }2.099 & 0.931 & \cellcolor[rgb]{ 0.82 0.93 0.53 }1.512 & 0.933 \\ 
\cmidrule{2-10}          & ($\chi$; $Q$) & \multicolumn{2}{c}{(0.417; 0.752)}   & \multicolumn{2}{c}{(0.603; 0.783)}   & \multicolumn{2}{c}{(0.411; 0.777)}   & \multicolumn{2}{c}{(0.742; 0.781)}   \\ 
\midrule 
\multirow[t]{1}[4]{3.0cm}{\centering Overall}  & ($\chi$; $Q$) & \multicolumn{2}{c}{(0.770; 12.988)}  & \multicolumn{2}{c}{(0.807; 13.309)} & \multicolumn{2}{c}{(0.763; 13.312)} & \multicolumn{2}{c}{(0.723; 13.326)} \\ 
\end{tabular} 
\end{ruledtabular} 
\centering
\includegraphics[width= 0.6\textwidth ]{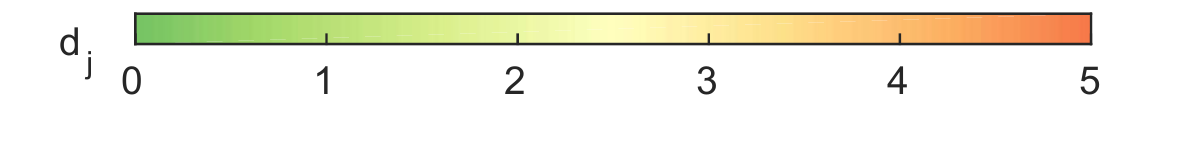}
\centering
\end{table*} 

\newpage

Fig.~\ref{fig:Compare_182574_DSS} shows the comparison of the Balmer line intensity profiles measured by DSS, where the DSS chords, see Fig.~\ref{fig:ExpSimu_polview} a), are labeled with increasing number from the bottom up to the X-point. The measured intensity in the forward field direction is systematically larger than the reversed field direction. For all three Balmer lines, the simulation successfully reproduces the profile shape, with the two peaks of the intensity, corresponding to the LFS target and the X-point/high-field-side (HFS) target. 
In the simulation, the location of the second peak (with the larger chord number) matches well with the experiment, while the first peak is shifted. 
Generally, the simulation underestimates the intensity, especially in the region between the two peaks, i.e., along the leg. At the first peak, the underestimation is small for ${D}_{5\rightarrow 2}$ and for ${D}_{6\rightarrow 2}$, while large for ${D}_{7\rightarrow 2}$. At the second peak, the value in the simulation is much smaller than that in the forward field measurement, while closer to that measured in the reversed field experiments. 

Fig.~\ref{fig:Compare_182574_pn} presents the experimental and simulation results for the neutral pressure in the two locations in the divertor. 
Both $p_{\text{div}}$ and $p_{\text{tmp}}$ in the forward field direction are higher than in the reversed field direction. The simulation result for $p_{\text{div}}$ in the standard approach is within the range of uncertainty of the forward field case, while the simulated $p_{\text{tmp}}$ matches within experimental uncertainty of the reversed field case. Compared to the full-field TCV SOLPS simulations studied in Ref.~\onlinecite{wensing_solps-iter_2021}, where the simulated $p_{\text{div}}$ systematically exceeded the measured value by a factor $\sim 400\%$, here in the reversed field, $p_{\text{div}}$ is overestimated only by $\sim 50\%$ at most. 

In summary, the validation in this case gives an overall agreement metric $\chi=0.770$ for the forward field direction, and $\chi=0.807$ for the reversed field direction. Good agreement is found for the $n_e$, $T_e$, $J_{sat}$, and $V_{pl}$ profiles at the outboard midplane and the $n_e$ and $T_e$ profile in the divertor entrance, for both field directions. $V_{fl}$ in the midplane and $p_\mathrm{div}$ in the forward field direction, and $p_\mathrm{tmp}$ in the reversed field direction also show good quantitative agreement. Poorer matches are found in $M_\parallel$ at the outboard midplane, $M_\parallel$ and $V_{fl}$ in the divertor volume, $q_\parallel$ at the low field side target, and $V_{pl}$ at the high field side target, in both field directions, and in $V_{pl}$ and $V_{fl}$ at the low field side target in the reversed field direction. 

\begin{figure}[htbp]
\includegraphics[width= 0.48\textwidth ]{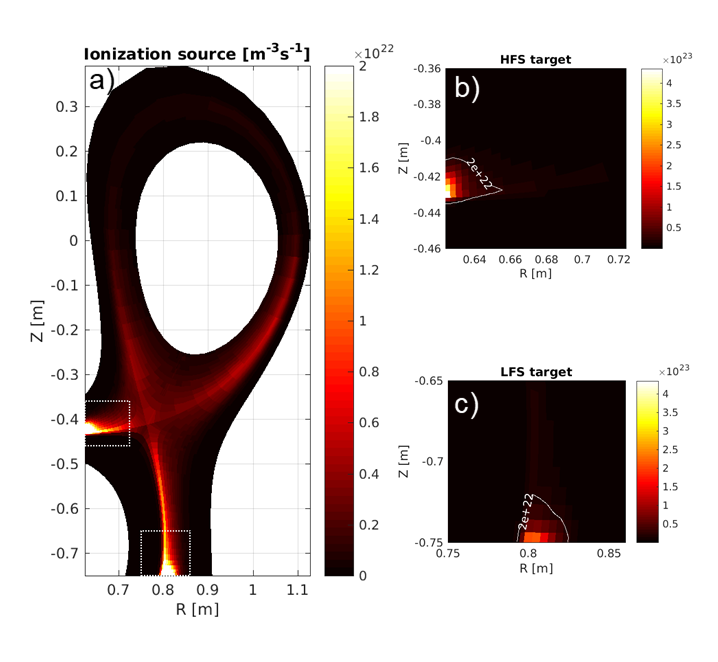}
\caption{\label{fig:Ionization_sub}\textbf{Simulated total ionization source.} In (a) the values at the two targets go up to $\sim 10^{23} \mathrm{m^{-3}s^{-1}}$, which is saturated in the color code, and are thus shown in the two sub figures (b) and (c).}
\end{figure}

To gain insight on role of the neutrals in the TCV-X21 scenario, we plot in Fig.~\ref{fig:Ionization_sub} the simulated total ionization sources (more precisely, what is shown is the source of electrons due to ionization with the generation of both $\mathrm{D}^+$ and $\mathrm{D_2}^+$). We can observe that in the SOL, most of the ionization happens along the separatrix, especially at the two targets. According to the SOLPS simulation, the ionization in the SOL (regions \raisebox{.5pt}{\textcircled{\raisebox{-.9pt} {2}}}, \raisebox{.5pt}{\textcircled{\raisebox{-.9pt} {3}}} and \raisebox{.5pt}{\textcircled{\raisebox{-.9pt} {4}}} in Fig.~\ref{fig:ExpSimu_polview}(b))  accounts for $\sim 65.0\%$ of the total ionization. The HFS divertor region (\raisebox{.5pt}{\textcircled{\raisebox{-.9pt} {3}}} in Fig.~\ref{fig:ExpSimu_polview}(b)) accounts for $\sim 14.9\%$ and the LFS divertor region (\raisebox{.5pt}{\textcircled{\raisebox{-.9pt} {2}}} in Fig.~\ref{fig:ExpSimu_polview}(b)) accounts for $\sim 31.2\%$. 
Another difference between these SOLPS-ITER simulations and the turbulence simulations in~\onlinecite{sales_de_oliveira_validation_2022} is the inclusion of impurity species. Here the SOLPS simulation includes carbon impurities. Their radiation is relatively weak in this scenario, $\sim15\%$ of the total input power. 

\subsection{\label{subsec:Gradient_method}Conjugate gradient method}

We explore here a systematic method to determine SOLPS-ITER input parameter values that lead to an overall improvement of the agreement metric $\chi$. This is done by minimizing the multi-variable function $\chi$ (we recall that $\chi=0$ indicates perfect experiment-simulation agreement: 
\begin{equation}
    \chi(\mathbf{x}), \quad
    \mathbf{x} = (\Gamma_\mathrm{D_2}, D_\perp, \chi_{e,\perp}),
\end{equation}
considering its dependence on the gas puff rate $\Gamma_{\mathrm{D_2}}$, the particle diffusivity $D_\perp$, and the electron thermal diffusivity $\chi_{e, \perp}$. As a proof of principle, only these three input parameters are used in this test, but one could, in principle, use all input parameters of the simulation subjected to tuning.

For this task, we apply the conjugate gradient method, an algorithm used to solve unconstrained optimization problems~\cite{hestenes_conjugate_methods_1952,press_numerical_2007}. The main advantage of this method, compared to the gradient descent method, is the avoidance of oscillating behaviors when calculating the gradient directions in the iterative minimization~\cite{press_numerical_2007}.The algorithm to determine the input parameters can be briefly described as follows: 

Step 1: The index $r$ indicates the iteration step number, with $r\in\{0,1,2,...\}$. In the first iteration, $r=0$, we compute the gradient $\mathbf{g}_0 = \nabla \chi(\mathbf{x_0})$ at the starting point $\mathbf{x}_0=(\Gamma_\mathrm{D_2,0}, D_{\perp,0}, \chi_{e,\perp,0})$. This is done using finite differences between $\chi(\mathbf{x_0})$ and three neighbouring simulations. Then, the minimization direction is set as $\mathbf{s}_0 = -\mathbf{g}_0$.

Step 2: For the current $r$, perform simulations along the direction $\mathbf{s}_r$ using the parameters determined as $\mathbf{x}_{r+1}'=\mathbf{x}_r + \lambda \mathbf{s}_r$, with the values of $\lambda$ chosen appropriately.

Step 3: Evaluate $\chi$ for all simulations in Step 2 and determine the local minimum, $\mathbf{x_{r+1}}$. 

Step 4: If an asymptotic convergence is not achieved, the scan continues using the new gradient $\mathbf{g}_{r+1} = \nabla \chi(\mathbf{x_{r+1}})$
, and the new scan direction (the conjugate gradient direction) is set by
\begin{equation}
    \mathbf{s}_{r+1} = -\mathbf{g}_{r+1}
    + \frac{g_{r+1}^2}{g_{r}^2}\mathbf{s}_r.
\end{equation} 

Step 5: Once $\mathbf{x}_{r+1}$ and $\mathbf{s}_{r+1}$ are determined in Step 3 and Step 4, restart from Step 2 with $r=r+1$ for the next iteration.

In this way, $\mathbf{x}_r$ is expected to converge to the minimum possible of $\mathbf{x}$, i.e., maximized agreement.

\begin{figure*}[htbp]
\includegraphics[width= 0.98\textwidth ]{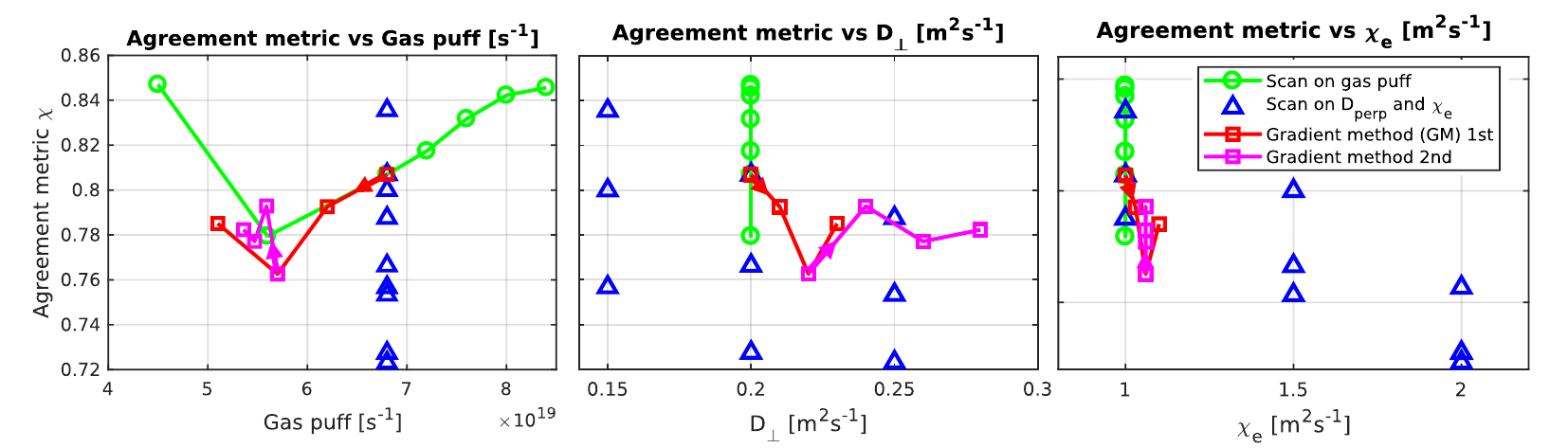}
\caption{\label{fig:Validation_grad_scan}
\textbf{Validation metric of simulations using the gradient method and parameter grid scan.} The $\chi$ of each simulation is plotted as a function of the gas puff $\Gamma_{\mathrm{D_2}}$, the particle diffusivity $D_\perp$, and the electron thermal diffusivity $\chi_{e, \perp}$. The gas puff scan of section ~\ref{subsec:Vali_Standard} is shown by the green circles, while the $D_\perp$ and $\chi_{e, \perp}$ scan is plotted as blue triangles. The simulations of the first and second iteration of the conjugate gradient method are plotted in red and magenta squares, respectively. The arrows denote the directions of searching.}
\end{figure*}

Fig.~\ref{fig:Validation_grad_scan} shows the results of performing two iterations of the conjugate gradient method. The agreement metric is computed with respect to the reversed field case. In this demonstration, the parameters were first normalized to their values at the starting point, such that all quantities have a similar order of magnitude, and then the conjugate gradient method was applied. Finally, the parameters were denormalized back to real values. In each step, in order to find an approximate minimum along $\mathbf{s}_{r}$, we performed simulations for three positive values of $\lambda$ and treated the one with minimum $\chi$ as the minimum point. The gradient needed to determine $\mathbf{s}$ was calculated numerically, using the metric differences between the point of interest and three neighboring points. 

The starting point was chosen from the simulation in Sec.~\ref{subsec:Vali_Standard}, with $\Gamma_\mathrm{D_2}=6.8\times 10^{19}\mathrm{m/s}$, $D_{\perp} = 0.2 \ \mathrm{m^2/s}$, and $\chi_{e,\perp}= 1.0 \ \mathrm{m^2/s}$. In the first iteration (Fig.~\ref{fig:Validation_grad_scan}, red line), the metric decreases first, then increases, revealing a minimum. We select this lowest point as the starting point of the second iteration (Fig.~\ref{fig:Validation_grad_scan}, magenta line), where we observe an alternating increase and decrease, without exhibiting a clear asymptotic convergence or improvement of the agreement metric. Overall, the best case (lowest $\chi$) is obtained in the first step, where $\Gamma_\mathrm{D_2}=5.7\times 10^{19}\mathrm{m/s}$, $ D_\perp=0.22\  \mathrm{m^2/s}$, and $\chi_{e,\perp}=1.06\  \mathrm{m^2/s}$. The procedure is stopped after these two iterations, as no improvement was achieved in the second step and due to the involved simulation costs (12 simulations were needed in total for these two minimization steps).

In Tab.~\ref{tab:validation_182574}, the $d_j$ and $\chi$ values for the simulations with the smallest overall metric $\chi$ found with the gradient method are presented in column "Gradient(Rev)". In general, the two step demonstration gives an overall improvement of the agreement by $5.5\%$. The majority ($21$ out of $32$) of the observables have been improved as indicated by the decrease of the corresponding $d_j$. Among them, $11$ observables for the gradient method show a decrease of the $d_j$ value larger than $10\%$. Several observables have been significantly improved (from disagreement to agreement), for example, the $n_e$ and $T_e$ profiles measured at the LFS target by the LPs, $T_e$ of RDPA, and $p_\mathrm{div}$. 

\subsection{\label{subsec:grid_scan}$D_\perp$ and $\chi_{e, \perp}$ scans}

As an alternative of the conjugate gradient method, we also conduct here a scan of $D_\perp$ and $\chi_{e, \perp}$, on the grids spanned by  $D_\perp=\{0.15,\  0.2,\  0.25\}\ \mathrm{m^2/s}$ and $\chi_{e, \perp}=\{1.0,\  1.5,\  2.0\}\ \mathrm{m^2/s}$, which are plotted as blue triangles in Fig.~\ref{fig:Validation_grad_scan}. For these nine simulations, the deuterium gas puff is fixed to be $\Gamma_\mathrm{D_2} = 6.8\times 10^{19} \ \mathrm{/s}$. In Fig.~\ref{fig:Validation_grad_scan}, we find that the simulation with the largest $D_\perp$ and $\chi_{e, \perp}$ value, $D_\perp=0.25\ \mathrm{m^2/s}$ and $\chi_{e, \perp}=2.0\ \mathrm{m^2/s}$ gives the best agreement.

In Tab.~\ref{tab:validation_182574}, the $d_j$ and $\chi$ values for the simulation with the smallest overall metric $\chi$ in the $D_\perp$ and $\chi_{e, \perp}$ scan are given in column "$D_\perp$  $\chi_{e, \perp}$ scan (Rev)". The overall increase of agreement given by $\chi$ is $\sim10.4\%$. $23$ out of $32$ observables have been improved as indicated by the decrease of their $d_j$.  Among them, $14$ observables show a decrease of their $d_j$ value by more than $10\%$. Several observables have been significantly improved (from disagreement to agreement), for example, the $T_e$ profile from LPs at the LFS and HFS targets, and $T_e$ from RDPA. 

\begin{figure}[htbp]
\includegraphics[width= 0.49\textwidth ]{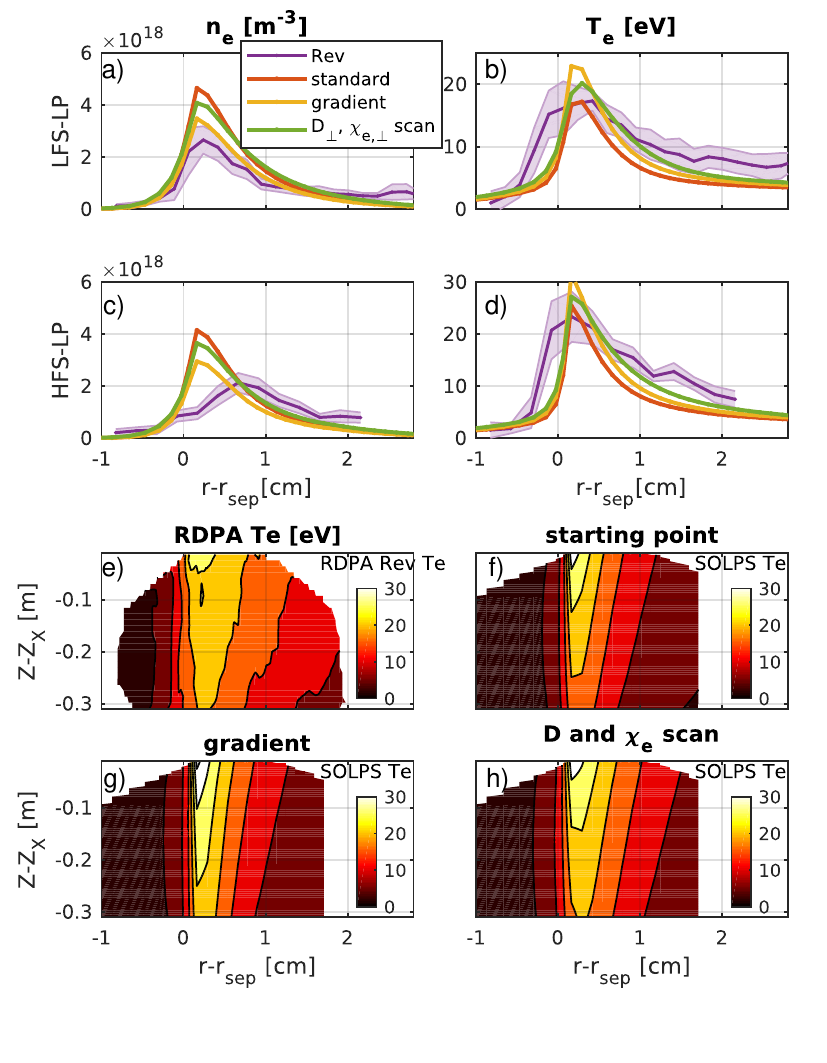}
\caption{\label{fig:Compare_file_improvement}
\textbf{Comparison of profiles from the standard simulation in Sec.~\ref{subsec:Vali_Standard} (starting point), the best case from gradient method, and the best case from the $D_\perp$ and $\chi_{e,\perp}$ scan.} We show the 1-D experimental profiles of reversed field (purple lines) and the SOLPS-ITER simulation profiles for (a) electron density and (b) electron temperature measured by the LPs at the low-field-side divertor target, and (c) electron density and (d) electron temperature measured by the LPs at the high-field-side divertor target. We also show (e) the 2-D profile of temperature measured by RDPA, and (f)-(h) the corresponding simulation results.}
\end{figure}

\section{\label{sec:Discussion}Discussion}

In Sec.~\ref{subsec:Vali_Standard}, a SOLPS-ITER simulation without drifts and with constant particle and energy diffusivity $D_\perp$ and $\chi_{e/i,\perp}$, was validated against the TCV-X21 reference case. The standard TCV L-mode values for $D_\perp$, $\chi_{e,\perp}$ and $\chi_{i,\perp}$ were used ~\cite{wensing_solps-iter_2021} and the gas puff rate was manually tuned to have a good upstream density match. The simulation shows good agreement with experimental data at the outboard midplane and the divertor entrance, with the exception of the parallel Mach number. The agreement is especially good for the electron density $n_e$ and temperature $T_e$. When approaching the divertor targets, a less good agreement is found, but the simulation captures the shape and order of magnitude of the $n_e$ and $T_e$ profiles. The quantitative match with the floating potential $V_{fl}$, plasma potential $V_{pl}$, parallel Mach number $M_\parallel$, and parallel heat flux $q_\parallel$, are not satisfactory. The large experimental-simulation distance for $q_\parallel$ can be mainly attributed to the small experimental uncertainties and a shift of the profile peak positions (Not shown). 
The mismatch of $V_{fl}$ and $V_{pl}$ with experimental data is mainly attributed to the omission of drifts in the simulations.

In Sec.~\ref{subsec:Gradient_method} and Sec.~\ref{subsec:grid_scan}, using two different methods, we obtained improved quantitative agreement compared with the standard approach in Sec.~\ref{subsec:Vali_Standard}. The best agreement case in the conjugate gradient method of Sec.~\ref{subsec:Gradient_method} features a decrease of gas puff, and slight increases in the two transport coefficients. From Fig.~\ref{fig:Compare_file_improvement}(a), we can observe that the significant improvement in the $d_j$ of the outer target $n_e$ is mainly due to the decrease of the peak value, as a result of a reduced gas puff. The improved match of $p_\mathrm{div}$ in the divertor region for the reversed field case is also related to the decrease of the gas puff, resulting in a decrease of neutral pressure (Fig.~\ref{fig:Compare_182574_pn}). The best agreement case in the $D_\perp$ and $\chi_{e, \perp}$ scan features a significant increase of the transport coefficients, which leads to an increase of the fall-off length of $n_e$ and $T_e$ at both targets and improves the match with the experiment profile. Similarly, $T_e$ in the divertor volume displays a shallower radial decrease, which agrees better with the RDPA measurements. This may indicate that the experimental case has a higher perpendicular transport compared to what is assumed in the simulation with the standard approach. 

\begin{figure}[htbp]
\includegraphics[width= 0.49\textwidth ]{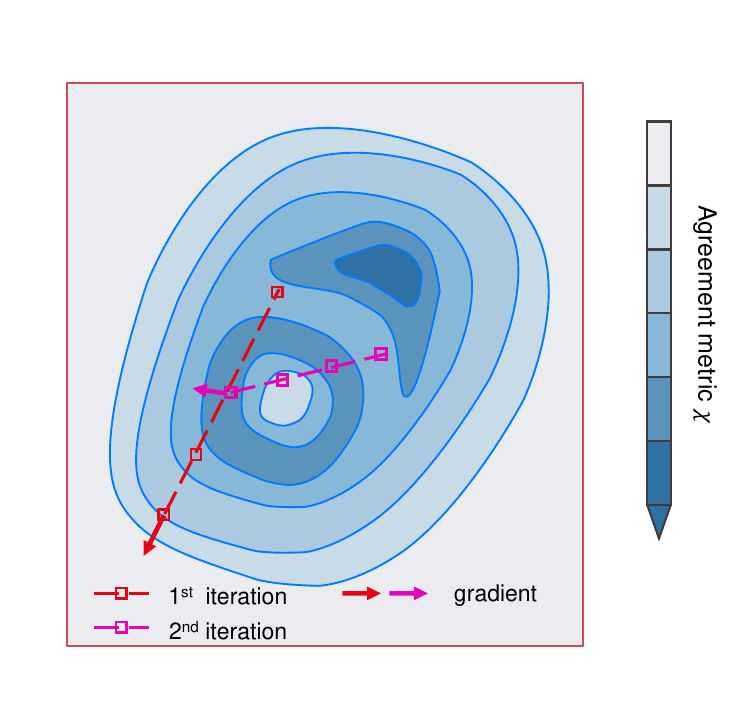}
\caption{\label{fig:Discussion_counter_example}\textbf{A simple example that could give a similar behavior for the conjugate gradient method as in our study.} For illustration, we consider a 2-D minimization problem with the filled contour plot. The first and second 1-D search are denoted by the red dash line and magenta dash line. The red arrow represents the direction of the first gradient (opposite to the first 1-D search direction), while the magenta arrow represents the second gradient. }
\end{figure}

The conjugate gradient method does not give the overall best agreement (lowest metric $\chi$) although it does improve the result marginally. This can be attributed to several possible factors. First, the conjugate gradient method does not guarantee the increase of the agreement level in every iteration step. As shown in an example in Fig.~\ref{fig:Discussion_counter_example}, which aims to reproduce a similar behavior as observed in Sec.~\ref{subsec:Gradient_method}, two iteration steps are not enough to reach the minimum point and non-monotonic behaviors can be found in the second iteration.  To go towards the minimum point, more iteration steps may be needed. 
Second, the finite difference method used for the gradient calculation might introduce non-negligible errors. In the calculation of the gradient in iteration 2, we assumed the gradient along the first direction of the 1-D search to be zero, which is also an approximation. Third, the step length of each 1-D search is limited by the numerical cost of each simulation. Therefore, the 1-D minimization can only be estimated approximately, with an error of the order of the step length. Possible solutions to these problems could be using other methods for the numerical differentiation, for example the central difference method, to get a better estimate of the gradient; or trying other minimization methods independent of the evaluation of the gradients, for example, the multi-dimensional simplex method~\cite{press_numerical_2007}. Compared to the improvement it brought, the conjugate gradient method tested here is found to be a computationally too expensive approach.

The ionization profile given by the SOLPS-ITER simulation (Fig.~\ref{fig:Ionization_sub}) is clearly different to what was assumed in the turbulence simulations in Ref.~\onlinecite{sales_de_oliveira_validation_2022}, where the ionization sources were assumed to be localized in the outer region of the confined plasma. This motivates further studies to add more realistic neutral particle profiles, or self-consistent inclusion of neutrals, to the turbulence simulations. 

\begin{figure}[htbp]
\includegraphics[width= 0.48\textwidth ]{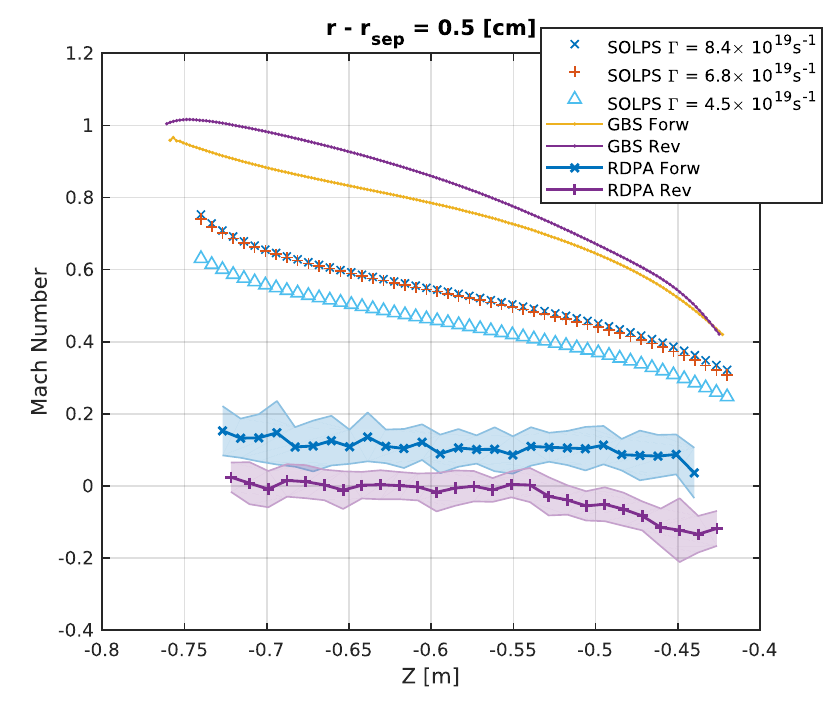}
\caption{\label{fig:Mach_1D}\textbf{The Mach number along the separatrix to the LFS target.} These 1-D plots consider the parallel Mach number along a flux surface (at $r-r_\mathrm{sep}=0.5 \mathrm{\ cm}$) in the SOL near the LFS separatrix. The RDPA measurements are plotted by solid lines with shaded errorbars. The SOLPS simulations from Sec.~\ref{subsec:Vali_Standard} with three gas puff values are given together with GBS simulations in Ref. ~\onlinecite{sales_de_oliveira_validation_2022}.}
\end{figure}

To explore the effect of neutrals in the flows in the divertor region, 
we plotted the parallel Mach number profile along a magnetic flux surface at $r-r_{\mathrm{sep}} =0.5 \mathrm{\ cm}$ in Fig.~\ref{fig:Mach_1D}. The GBS turbulence simulation without neutrals from Ref.~\onlinecite{sales_de_oliveira_validation_2022} and the SOLPS simulation results are of the same order of magnitude. 
The simulated parallel flows point towards the target (positive Mach number) and reveal a significant flow all along the divertor leg, being somewhat weaker in SOLPS-ITER. Instead, the Mach number measured with RDPA is much small, close to zero. Both GBS and SOLPS simulations show an increasing Mach number as approaching the divertor target, while the RDPA measurements feature a flat profile. Comparing the GBS simulations in reversed and forward toroidal field directions, we find that the effect of drifts is small compared with the difference between simulation and RDPA measurement. Ref.~\onlinecite{sales_de_oliveira_validation_2022} suggested the difference between the GBS simulations and RDPA measurements to be due to the ionization source along the divertor leg and primarily to potentially be located just in front of the target. In this study using the SOLPS simulation with neutrals, the Mach number is lower than the GBS, but the flows in the simulations are still significant along the entire divertor leg and considerably higher than in the experiment. This raises questions in the Mach number measurements with RDPA in these conditions or the model used for its interpretation. 
Further investigation is needed in order to disentangle the differences between the flow velocities presented here. 

\section{\label{sec:Conclusion}Conclusion}
SOLPS-ITER transport fluid simulations without drifts and with uniform particle and energy cross-field transport coefficients were qualitatively compared and quantitatively validated against the TCV-X21 experimental reference case from Ref.~\onlinecite{sales_de_oliveira_validation_2022}. Three Balmer lines measured across the outer divertor leg by DSS and two divertor neutral pressure measurements from the Baratron gauges were added to the publicly available TCV-X21 dataset. In the standard approach, where SOLPS-ITER input parameters are tuned to match upstream quantities, qualitative comparisons of profiles from upstream to the divertor were carried out. 
As expected, in this standard approach, agreement between simulation and experiment is found in the outer midplane and at the divertor entrance, with the exception of parallel Mach flows. Reduced agreement is found in the divertor volume and at the divertor targets, for quantities such as density, temperature, and plasma potential profiles.
Despite TCV-X21 being a (near) sheath-limited plasma designed to minimize the effect of neutrals in the SOL/divertors, the simulation still finds a $\sim 65.0\%$ of the ionization happening in the SOL, with $\sim 14.9\%$ in the HFS divertor region and $\sim 31.2\%$ in the LFS divertor region. The simulation also shows $\sim 15\%$ of the input power to be radiated by carbon impurities. Using the quantitative validation metric $\chi$, in a proof-of-principle test, we use the conjugate gradient method and $D_\perp$ and $\chi_{e, \perp}$ scans to improve the agreement level, resulting in a $5.5\%$ and $10.4\%$ improvement respectively compared to the results achieved using the standard approach. This suggests that a reduced gas puff and increased particle and energy transport coefficients compared to what is used when exclusively trying to match upstream profiles lead to a better match with the experimental case, mainly via decreasing the peak target density and broadening the density and temperature profiles.  While the performance achieved here with the conjugate gradient method is rather modest, this method may be improved by using finer steps and better numerical differentiation methods. Other algorithms, such as multidimensional simplex method may, however, be better a solution for the problems presented here in the iterative method to determine the input parameters resulting in an optimal match with the experiment.

The SOLPS simulations with neutrals show a significant portion of neutral ionization to occur in the SOL, a major difference compared with the assumption used in the first turbulence code validation in the TCV-X21 validation case in Ref.~\onlinecite{sales_de_oliveira_validation_2022}. The parallel flows in the divertor observed in SOLPS and GBS turbulence simulations from Ref.~\onlinecite{sales_de_oliveira_validation_2022} are similar in shape, with the GBS divertor flows systematically larger in comparison to the SOLPS flows. This suggests some flow reduction in the divertor by the neutrals. The parallel Mach numbers from SOLPS-ITER are, however, still substantially larger than those measured with RDPA, raising questions on the latter that will be further explored in the future. \\ 
The results in this work provide useful information for future turbulence simulations of TCV with neutrals,  while suggesting that the contribution of the neutrals to the flow velocity and, therefore, to the parallel heat flow towards the targets, should be further investigated.  
\section{\label{sec:Acknowledgement}Acknowledgement}
This work has been carried out within the framework of the EUROfusion Consortium, partially funded by the European Union via the Euratom Research and Training Programme (Grant Agreement No 101052200 — EUROfusion). The Swiss contribution to this work has been funded by the Swiss State Secretariat for Education, Research and Innovation (SERI). Views and opinions expressed are however those of the author(s) only and do not necessarily reflect those of the European Union, the European Commission or SERI. Neither the European Union nor the European Commission nor SERI can be held responsible for them. This work was supported in part by the Swiss National Science Foundation.
\appendix
\section{{\label{appendix:vali_method}}Definition of the quantities in the validation methodology}
This appendix summarizes the quantities used in the validation procedure, following the same method as in Ref.~\onlinecite{sales_de_oliveira_validation_2022}. 

The sensitivity $S_j$, for the observable $j$, is given by
\begin{equation}
    S_{j}=\exp \left(-\frac{\sum_{i} \Delta e_{j, i}+\sum_{i} \Delta s_{j, i}}{\sum_{i}\left|e_{j, i}\right|+\sum_{i}\left|s_{j, i}\right|}\right),
\end{equation}
where $e_{j,i}$, $\Delta e_{j,i}$,  $s_{j,i}$, and $\Delta s_{j,i}$ denote, respectively, the experimental values, their uncertainties, the simulation values, and their uncertainties, defined on a series of discrete data points $i\in\{1, 2,  ..., N_j \}$.

The hierarchy weighting $H_j$ is defined as 
\begin{equation}
    H_j=(h_{sim}+h_{exp}-1)^{-1},
\end{equation}
where $h_{sim}$ and $h_{exp}$ are the simulation and experimental primacy hierarchy level for an observable, being higher the higher number of assumptions and/or measurement combinations used to obtain the observable.

The overall agreement metric $\chi$ and the quality $Q$ of a set of observables are obtained by: 
\begin{equation}
    \chi=\frac{\sum_{j} R_j\left(d_{j}\right) H_{j} S_{j}}{\sum_{j} H_{j} S_{j}}
    ,\qquad
    Q=\sum_j H_j S_j,
\end{equation}
where \textit{the level-of-agreement function} $R_j(d_j)$ is an increasing function of the normalised simulation-experimental distance $d_j$ (eq.~\ref{eq:Vali_d_J}), defined as
\begin{equation}
    R_j\left(d_{j}\right)=\frac{\tanh \left[\left(d_{j}-1 / d_{j}-d_{0}\right) / \lambda\right]+1}{2}.
\end{equation}
In this work we set $d_0=1$, $\lambda=0.5$, as in Ref.~\onlinecite{ricci_approaching_2015}. $R_j$ takes values between $0$ and $1$. It is used to unify the distance to a level of agreement with fixed range, from perfect agreement $(0)$ to complete disagreement within errorbars $(1)$.

\bibliography{Bib}

\end{document}